\documentstyle[12pt]{article}
\input epsf
\def\oneh{\frac{1}{2}}
\def\x{\mbox{\boldmath $x$}}
\def\w{{\bf w}}
\def\ws{{\bf w}}
\def\wt{{\bf w^*}}

\def\dwt{d\mu(\wt)}
\newcommand{\beq}{\begin{equation}}
\newcommand{\eeq}{\end{equation}}
\newcommand{\beqa}{\begin{eqnarray}}
\newcommand{\eeqa}{\end{eqnarray}}
\newcommand{\beqas}{\begin{eqnarray*}}
\newcommand{\eeqas}{\end{eqnarray*}}

\begin{document}

\title {Retarded Learning: Rigorous Results from Statistical Mechanics}
\author{Didier Herschkowitz$\dagger$ \and Manfred Opper$\ddagger$ \and \\
$\dagger$ Laboratoire de Physique Statistique de L'E.N.S.,
Ecole Normale Sup\'erieure, Paris, France \\
$\ddagger$ Neural Computing Research Group, Aston University,
United Kingdom
}
\maketitle
\begin{abstract}
We study learning of probability distributions 
characterized by an unknown symmetry direction. 
Based on an entropic performance measure and the
variational method of statistical mechanics we
develop exact upper and lower bounds on the scaled critical
number of examples below which learning of the
direction is impossible. 
The asymptotic tightness of the bounds suggests
an asymptotically optimal method for learning nonsmooth
distributions.
\end{abstract}
PACS numbers: 87.10.+e, 05.20.+m, 02.50.-r

In recent years, methods of Statistical Physics have 
contributed important insights into the theory of learning 
with neural networks and other learning machines 
(see e.g. \cite{Seungandall92,Waal93,OpKi96}).
Among the most prominent discoveries of statistical mechanics in this 
field is the occurence of phase transitions in the progress of learning, 
when the number of presented example data is gradually increased and 
the dimensionality of data and model parameter is large. 

Besides the ubiquity of phase transitions in discrete parameter
models, they are typically observed when the learning problem
contains symmetries 
which are spontaneously 
broken when the scaled number of examples increases beyond a critical value
\cite{Seungandall92,Schwarzeandall92,Opper94}.
Although phase transitions in neural networks have been analysed
extensively by the method of replicas \cite{Mezzardandall87} 
it is usually hard to present a rigorous analysis (for an exception see
e.g. \cite{HKST96} and the recent attempts of
Talagrand \cite{Talagrand}). 
Hence, this often precludes a digestion of the 
interesting results by researchers outside the community of 
statistical physicists working on disordered systems.
Unfortunately, also other standard techniques
based on asymptotic expansions \cite{AmariMurata93} 
will not apply in these cases. They are only valid
when the number of data is much larger than the number of parameters. 

In this letter we will present a rigorous and simple
approach to these problems.
We combine information theoretic bounds for
the performance of statistical estimators (see e.g. 
\cite{HausslerOpper97,bnFisher,HerschkowitzNadal99})
with an elementary variational principle of statistical
physics \cite{FeynmanHibbs}. This will allow us 
to compute rigorous upper and lower bounds for the critical number of examples 
at which a transition occurs. 

We will explain our method for the case of 
retarded unsupervised learning which has been 
analysed before using the replica framework
(see e.g. \cite{BiehlMietzner93,ReimannVdBroeck,BuhotGordon98,Schottky95}).
The goal of unsupervised learning is to 
find a nontrivial structure in a set of data which reflects
the properties of the underlying data generating mechanism rather
than being an artefact of statistical fluctuations.
The phenomenon of retarded learning describes 
the fact that for some high dimensional 
probability distributions, it is impossible at all
to predict the underlying structure (usually a symmetry
axis) if the (scaled) 
number of data is below a certain critical value. 
Only above this value, estimation of the structure can start.

We adopt a probabilistic, Bayesian formulation of unsupervised
learning following 
\cite{OpperHaussler95,HausslerOpper97,ReimannVdBroeck}. 
We model a situation where the probability distribution of 
the data is characterized by a single unknown 
rotational-symmetry direction $\wt$.
More specifically, we assume that a set of $t$ data 
$\x^{t}=\x_1,\ldots,\x_{t}$,
has been generated independently by $t$ samplings from a distribution 
of the form
\beq
P(\x|\wt)= P_0(\x)
\;\exp\left(-V(\lambda)\right)
\label{probab}
\eeq
where $P_0(\x)=\frac{1}{(2\pi)^{N/2}} \exp
(-\x^2 /2)$ is a spherical Gaussian distribution and
$\lambda\doteq \wt\cdot\x$ is the projection of the N-dimensional
data vector $\x$
on the direction defined by $\wt$. 
The distribution of the projection is given by 
$p(\lambda)=\frac{1}{\sqrt{2\pi}}\exp(-\lambda^2 /2 - V(\lambda))$.
In the following, averaged quantities with 
respect to $p(\lambda)$ will be denoted with
an overline 
$\overline{(..)}= \int d\lambda p(\lambda) (..)$.  
Based on the set of data $\x^{t}$, 
the goal of a learner is to produce an estimate
$\hat{P}_{t}(\x|\x^{t})$ for the true distribution $P(\x|\wt)$. 
$\hat{P}_{t}$ will not necessarily belong
to the given parametric class (\ref{probab}).
Our approach relies very much on the choice of a specific
measure for the quality of the estimation. 
Rather than computing the overlap between an estimated direction
and the true $\wt$, we choose a quantity which is directly measuring 
our ability to compress the data based on the information we have 
gained on the structure of $P$. This is related to the averaged 
{\em relative entropy} (Kullback Leibler (KL) divergence) 
between the true distribution and the estimate
\beq\label{averageKL}
L(\hat{P}_t, \wt) \doteq  
\int d\x^{t} P(\x^t|\wt)
\int d\x P(\x|\wt)\ln\frac{P(\x|\wt)}{\hat{P}_{t}(\x|\x^{t})}
\eeq
where $P(\x^t|\wt)$ is shorthand for the product distribution
$\prod_{i=1}^t P(\x_i|\wt)$ and $d\x^{t} = \prod_{i=1}^t d\x_i$.
We will further adopt a Bayesian approach where we assume 
that "nature" draws the true parameter $\wt$ at random from a 
(noninformative) prior distribution $p(\wt)$ and 
associate measure $\dwt=p(\wt)d\wt$ 
given by the uniform distribution on the sphere with radius $\|\wt\|^2=1$.
The case of a discrete prior will be discussed later.

The progress of the learning will be measured by the cumulative risk defined by 
\beq\label{cumrisk}
R_m(\hat{P})=\sum_{t=0}^{m-1} \int \dwt L(\hat{P}_t, \wt)
\eeq
This measure of loss has a variety of important applications in 
information theory, game theory and mathematical finance (see e.g.
\cite{Cover,HausslerOpper97}). E.g., it is proportional to the  
expected extra number of bits (assuming a reasonable quantization of the $\x_i$) 
we have to suffer in 
compressing the observed data when their distribution is unknown and a sequential 
estimate is used instead \cite{Cover}. As we will see in a moment it has also an 
important meaning in statistical physics.

A first attempt to study retarded learning by using bounds
on (\ref{cumrisk}) was undertaken by \cite{HerschkowitzNadal99}.
However the bounds were too weak to give a nonzero
bound on the critical number of examples below which learning is impossible. 

An elementary calculation shows that the posterior probability
$
\hat{P}^{Bayes}_{t}(\x|\x^t)=
\frac{\int d\mu (\w) \ P(\x^t|\w) P(\x|\w)}
{\int d\mu (\w')\ P(\x^t|\w')}
$
achieves
the {\em minimum} risk $R_m^{Bayes}=R_m(\hat{P}^{Bayes})$ 
over all choices of estimators. 
Inserting this estimator into (\ref{cumrisk}) and using (\ref{probab})
we get
\beqa\label{Bayloss}
R^{Bayes}_m & = & 
\int d\mu (\wt)\int d\x^m P(\x^m|\wt) \left[-\ln\ \int  d\mu (\ws)
e^{-\sum_i \left\{V(\ws\cdot\x_i)- V(\wt\cdot\x_i)\right\}}\right]
\eeqa
The last line looks very much like an averaged free energy
in statistical mechanics for a system with hamiltonian
$\sum_i \left\{V(\ws\cdot\x_i)- V(\wt\cdot\x_i)\right\}$.
Hence we can expect that useful bounds for this quantity can be derived 
using the standard variational principle of 
statistical mechanics \cite{FeynmanHibbs} for the free energy
\beqa\label{varia}
-\ln \int d\mu(\w) e^{-H(\w)} \leq -\ln \int d\mu(\w) e^{-H_0(\w)} 
 + \left<H-H_0\right>_0
\eeqa
which bounds the free energy of a system with hamiltonian $H$
in terms of the free energy of a trial hamiltonian $H_0$ plus a correction
term. The brackets 
$\left<(..)\right>_0=
\int d\mu (\w) e^{-H_0(\w)}(..)/\int d\mu (\w') e^{-H_0(\w')}$ 
denote an average with respect to the Gibbs
distribution defined by $H_0$. Using appropriate choices for
$H$ and $H_0$, we will get both upper and lower bounds on 
(\ref{Bayloss}). 

We begin with the lower bound. We set
$H_0 =\sum_i \left\{V(\ws\cdot\x_i)- V(\wt\cdot\x_i)\right\}$ 
and $H= \sum_i \left\{\lambda V(\ws\cdot\x_i)- \gamma V(\wt\cdot\x_i)\right\}$ 
where $\lambda,\gamma>0$ are variational parameters. 
Averaging both sides of (\ref{varia}) over $P(\x^m|\wt)$ and $p(\wt)$
using Jensen's inequality
in the second line, we derive the lower bound on $R_m^{Bayes}$
\beqa\label{theorem1}
\nonumber
R_m^{Bayes} & \geq & 
\int d\mu(\wt) d\x^m P(\x^m|\wt)
\left[-\ln \int d\mu(\w) e^{-H}\right] +m(\gamma-\lambda)\overline{V}  \\
  & \geq & 
-\int d\mu(\wt)\ln \int d\mu(\w) 
\left[\int d\x P_0(\x)
\frac{e^{-\lambda V(\x \cdot \w)}}{e^{(1-\gamma) V(\x \cdot \wt)}}\right]^m 
+m (\gamma-\lambda)\overline{V}  \\ 
  & = &
-\ln\left\{\int_{-1}^1 dq\; W_N(q) 
\left[F_{\lambda\gamma}(q)\right]^m\right\} +m\;(\gamma-\lambda) \overline{V}   
\nonumber
\eeqa
where 
$W_N(q) = \int d\mu(\ws)  \; \delta(q-\ws.\wt) \propto
(1-q^2)^{\frac{N-3}{2}} $ and
$$
F_{\lambda\gamma}(q)= \int Dx \int Dy\; 
e^{-\lambda V(x) - (1 - \gamma) V(xq+y\sqrt{1-q^2})},
$$
with the gaussian measure
$Dx=e^{-\oneh x^2}dx/\sqrt{2\pi}$. 
This bound holds for every $N$ and every $m$. 

\begin{figure}[h]
\centerline{\epsfxsize=220pt\epsfbox{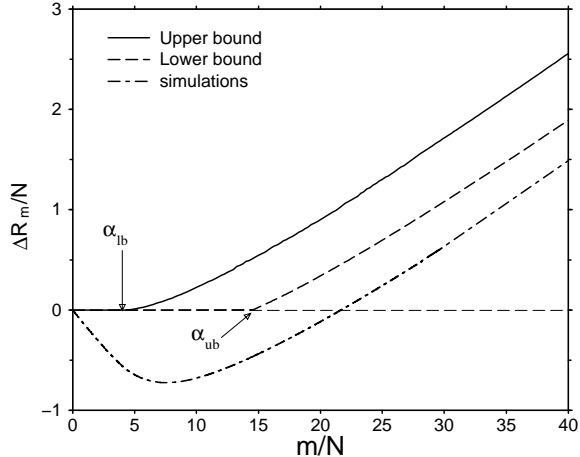}}
\caption{Upper and lower bound for the Bayes Risk 
$\Delta R_m /N=(R_m^{Bayes}-R_m^{trivial})/N$ in
the limit $N\to\infty$ for the gaussian case $A=-0.5$. $\alpha_{lb}=4$.  
Simulations are for the plugin estimator and show 
$(R_m(P^{triv}) - R_m(\hat{P}^p))/N$ for $N=100$  
averaged over 50 data sets.
}
\label{fig}
\end{figure}

To show the phenomenon of retarded learning
we will compare the cumulative risk of the Bayes 
estimator to the risk of a {\em trivial estimator} which assumes that
there is no specific structure in the data and always predicts with
the spherical distribution $\hat{P}_t^{triv}(\x)=P_0(\x)$ thereby
achieving the trivial total risk
$
R_m^{triv} =
R_m(\hat{P}^{triv}) =-m\overline{V} 
$.
Note, that
$\overline{V}$ is a nonpositive quantity.
We are interested in the difference 
$\Delta R_m=R_m^{triv}-R_m^{Bayes}$ between the trivial risk and the Bayes risk.
Taking the thermodynamic limit
$N\to\infty$ with $\alpha=m/N$ fixed, we can evaluate the integral 
in (\ref{theorem1}) by Laplace's method which gives the asymptotic
upper bound for $\Delta R_m$
\beqa\label{theorem3}
\limsup_{N\to\infty}\Delta R_{\alpha N}/N & \leq &
\min_q\left\{\frac{1}{2} \ln(1-q^2) + \alpha \ln  F_{\gamma\lambda}(q)\right\} 
    + \alpha (\lambda-\gamma-1) \overline{V}  
\eeqa   
For sufficiently small 
$\alpha$, the bound (\ref{theorem3}) is optimized
for $\gamma=0$ and $\lambda=1$.
For any potential $V$ having the 
property $\overline{\lambda}=0$ (ie. when the problem is not trivially
learnable by computing the mean of the data) 
there is a critical value 
$\alpha_{lb}=(1-\overline{\lambda^2})^{-2}$ 
such that as long as $\alpha\leq \alpha_{lb}$ the minimizer is $q=0$ 
and 
$\lim_{N\to\infty} \Delta R_{\alpha N} / N \; \leq 0$ 
(see Fig. \ref{fig}).
Since the Bayes risk is minimal, we have $\Delta R_m\geq 0$ and we conclude that
$
\lim_{N\to\infty} \Delta R_{\alpha N} / N\;= 0
$
at least for $\alpha\leq \alpha_{lb}$. 
This proves the existence of a region of retarded learning, where even 
the risk of the optimal Bayes estimator is 
to leading order in $N$ as large as the risk 
of a trivial estimator which assumes that there is no any 
spatial structure at all. The bound $\alpha_{lb}$ agrees
with the critical $\alpha$ obtained in the replica analysis of
\cite{ReimannVdBroeck}.

We next derive a lower bound on $\Delta R_m$.
Using the fact that $R^{Bayes}_m$ is the minimimum cumulative risk
over any choice of estimators,   
for any distribution $Q(\x|\w)$ and estimator 
$
\hat{Q}(\x|\x^t)=\frac{\int d\mu (\w) \ Q(\x^t|\w) Q(\x|\w)}
{\int d\mu (\w')\ Q(\x^t|\w')}
$ 
we have $R_m^{Bayes}\leq R_m(\hat{Q})$.
Setting $H = -\ln\frac{Q(\x^m|\w)}{P(\x^m|\wt)}$ in (\ref{varia})
and restricting ourselves to the class of trial Hamiltonians $H_0$ which do not
depend on $\x^m$, it can be shown that the optimal choice 
is the data average $H_0 = \int d\x^m\; P(\x^m|\wt)\;H$,
for which, on average, the correction term in (\ref{varia})
vanishes. This yields
\beqa\label{upperdemo}
   R^{Bayes}_m  & \leq  &   R_m(\hat{Q})  \\  
\nonumber  
  & = &
-\int d\mu (\wt) d\x^m P(\x^m|\wt)\ln\frac{\int d\mu (\ws)\ Q(\x^m|\ws)}
{P(\x^m|\wt)}  \\ 
\nonumber
  & \leq &   
-\int d\mu (\wt) \ln \int d\mu (\ws)  
\exp\left[ -m \int d\x P(\x|\wt)
\ln\frac{Q(\x|\ws)}{P(\x|\wt)}\right] 
\eeqa
We now have to find a good choice for $Q(\x|\w)$. For 
$Q(\x|\w)$ with a structure of the form
(\ref{probab}), we set $Q(\x|\w)= P_0(\x)
\;\exp\left(-U(\w\cdot\x)\right)$. In the thermodynamic limit
we get the lower bound for $\Delta R_m$
\beqa\label{theorem5}
\liminf_{N\to\infty}\Delta R_{\alpha N}/N \geq  
\max_q \left\{\frac{1}{2} \ln(1-q^2) \right.  
   \left. - \alpha \int Dx \exp(-U_q(x))U(x) \right\}
\eeqa   
with
\beq\label{potopt}
U_q(x)=
-\ln\int Dy \exp \left[- V(xq+y\sqrt{1-q^2})\right].
\eeq
It is easy to see that for any $q$, the expression in 
the curly brackets
of (\ref{theorem5}) is maximised for $U(x)=U_q(x)$.
With this choice for $U$, we find that there exists
an $\alpha_{ub}$ such that for $\alpha>\alpha_{ub}$, we have 
$\lim_{N\to\infty}\Delta R_{\alpha N} / N\;> 0$
which means that now the performance of the Bayes risk  
is better than the trivial risk and
a nontrivial estimation of the direction
$\wt$ is possible (see Fig. \ref{fig}). $\alpha_{ub}$ gives an upper 
bound on the region of retarded learning 
but has no simple analytical expression. 

Our approach is also easily applied to a
discrete prior distribution, e.g. a uniform distribution 
on the hypercube \cite{Coppelliandall99}.  
Again, for small $\alpha$ we find a region of
retarded learning $\Delta R_m^{Bayes}/N=0$ at least for 
$\alpha\leq \alpha_{lb}$ where $\alpha_{lb}$ is 
exactly the same as for the spherical prior. 

For illustration, we apply the bounds to the simple case of a 
Gaussian distribution
for which the integrals can be done analytically. Other
distributions will be discussed in \cite{Herschko}.
We set
$
P(\x|\wt)= \frac{1}{({2\pi})^{N/2}(1+A)}
\exp\left(-\frac{\x^2}{2} + \frac{A}{2(1+A)} (\x\cdot\wt)^2\right).
$
The data are
normally distributed with unit variance in all directions perpendicular
to $\wt$ and with variance $1+A$ in the direction $\wt$.
The upper and lower bounds (\ref{theorem3}) and (\ref{theorem5})
(optmized w.r.t. $\lambda$
and $\gamma$) are shown 
in Figure \ref{fig} for $A=-0.5$ for which we obtain $\alpha_{lb}=1/A^2=4$.
We have compared the bounds with numerical simulations. Since it is
hard to compute the Bayes optimal estimation algorithmically, 
we have used the following (suboptimal) algorithm instead. 
We have computed the direction $\hat{\w}(\x^t)$ for $\wt$ which
maximises the posterior probability
of the data. The estimate of the distribution is given by the plugin estimator
$\hat{P}^p_t(\x|\x^t) \doteq P(\x|\hat{\w}(\x^t))$ which has 
the KL divergence 
\beqas
\int d\x P(\x|\wt) \ln\frac{P(\x|\wt)}{\hat{P}^p_t(\x|\x^t)}
=\frac{A^2}{1+A}(1-(\wt\cdot\hat{\w}(\x^t))^2)
\eeqas
and the cumulative entropic risk $R_m(\hat{P}^p)$ 
can be easily approximated numerically by averaging over a large 
number of data sets. 
Figure \ref{fig} shows the difference
$R_m(P^{triv}) - R_m(\hat{P}^p)$. Since the Bayes risk is minimal, 
the upper bound on $\Delta R_m$ is also an upper bound on every estimator
while the lower bound is only a lower bound on the Bayes risk.
We see that until $\alpha\approx 4$, $\Delta R_m$ is negative and decreases
linearly. Since the slope of the dash-dotted curve is proportional
to the (relative) {\em instantaneous} loss (i.e. the non cumulative risk), 
the plugin estimator
is in the retarded learning regime, its instantaneous
loss is even bigger than that of the trivial estimator. 
This is due to the fact  
that the plugin estimator has to keep the elliptic
form of the distribution $P(\x|\w)$ which is always different from 
spherical when $\hat{\w}(\x^t)\neq 0$. 
Between $\alpha\ge 4$ and $\alpha \le 7.5$, the performance 
of the plug-in estimator improves, the slope increases but is still negative.
For $\alpha >7.5$, 
the performance of the plugin estimator is better than the trivial one and
the curve starts to increase.  
The Bayes estimator does not have this kind of disadvantage 
and can take a form closer to the spherical distribution 
in the retarded learning region by smoothing over the parameters
$\wt$. 

Both bounds in figure \ref{fig} show the same type
of asymptotic growth for $\alpha\to\infty$. 
The asymptotics can be found analytically
by expanding both bounds (\ref{theorem3}) and (\ref{theorem5})
for $q\to 1$. For a smooth potential $V$
both bounds give asymptotically the same logarithmic growth 
$R_m^{Bayes}/N \simeq 1/2\ln\;\alpha$ which
can also be obtained by well known
asymptotic expansions involving the {\em Fisher information matrix}
\cite{ClarkeBarron90,rissanen96,bnFisher}.
On the other hand, our bounds can also be used
when these standard asymptotic expansions do not apply, e.g.
when the potential exhibits a discontinuity \cite{ReimannVdBroeck}
of the form 
$V(\lambda)=-\ln\:2\Theta(\lambda)+U(\lambda)$ 
with corresponding projection distribution 
$p(\lambda)=\Theta(\lambda) 
\frac{2}{\sqrt{2\pi}}\exp(-\lambda^2 /2 - U(\lambda))$ 
where $\Theta(\lambda)$ is the Heaviside function and $U$ is smooth,
Our bounds yield the asymptotic scaling $R_m^{Bayes}/N \simeq \ln\;\alpha$. 

The asymptotic matching of our bounds has an important
consequence for computing asymptotically good approximations
to Bayesian predictions. Such approximations are easily derived
for smooth potentials by local expansions
of the posterior distribution around its maximum (\cite{Berger}). However, 
this technique will obviously fail for nonsmooth potentials. On the other hand, 
our results show that the estimate $\hat{Q}$ in 
(\ref{upperdemo}) which uses the {\em smooth} optimizing potential 
$U_q$ (\ref{potopt}), has the same asymptotic performance as the Bayes optimal 
estimate $R_m^{Bayes}$. The smoothness of $U_q$ will again enable 
local expansions. For example, following (\ref{potopt}), the case 
$V(\lambda)=-\ln\:2\Theta(\lambda)$ can be well estimated
using $U_q(\lambda)=-\ln\:2H(-q\lambda/\sqrt{1-q^2})$ 
with $H(x)=\int_x^\infty Dx$
and where the maximizer $q$ of the right hand side
of (\ref{theorem5}) is a function of $\alpha$.

In this letter, we have put the phenomenon
of retarded learning first established by 
the replica method on a rigorous footing. 
Our method relies on a general information theoretic performance measure
for learning probability distributions which is related
to the free energy of statistical physics. A variational principle yields
a controlled approximation to this quantity by providing exact
upper and lower bounds which are valid 
for arbitrary dimensionality of the problem. We expect that this 
framework is flexible enough to be
applicable to more complex and realistic probabilistic models.
It may also be useful for constructing criteria that help to decide 
if structures estimated from a dataset in a high dimensional
space reflect a real feature of the underlying data generating mechanism
or if the result is expected to be a spurious effect of random fluctuations.

We are grateful to J.-P. Nadal for helpful discussions.

\end{document}